# Deformation insensitive thermal conductance of the designed Si metamaterial

Lina Yang[1], Quan Zhang[2], Gengkai Hu[1], Nuo Yang[3*]

[1] School of Aerospace Engineering, Beijing Institute of Technology, Beijing 100081, China

[2] School of Mathematical and Statistical Sciences, University of Galway, University Road, Galway, H91 TK33, Ireland

[3] School of Energy and Power Engineering, Huazhong University of Science and Technology, Wuhan 430074, China

Email: nuo@hust.edu.cn (N. Yang)

**Abstract**

The thermal management have been widely focused due to broad applications. Generally, the deformation can largely tune the thermal transport. The main challenge of flexible electronics/ materials is to maintain thermal conductance under large deformation. This work investigates the thermal conductance of a nano-designed Si metamaterial constructed with curved nanobeams by molecular dynamics simulation. Interestingly, it shows that the thermal conductance of the nano-designed Si metamaterial is insensitive under a large deformation (strain~-41%). The new feature comes from the designed curved nanobeams which makes a quasi-zero stiffness. Further calculations show that, when under a large deformation, the average stress in nanobeam is ultra-small (<151 MPa) and its phonon density of states are little changed. This work provides valuable insights on multifunction, such as both stable thermal and mechanical properties, of nano-designed metamaterials.

**Keywords:** Silicon metamaterial, deformation effect, thermal conductance, stress and strain, flexible electronics

## 1. Introduction

Rationally designed metamaterials by the advanced fabrication techniques[1,2] have attracted great attention due to their new functionality, such as high strength/stiffness to weight ratio[3], recoverability under strain[2], low thermal conductivity[4-8], damage resistance[9] and quasi-zero-stiffness[10]. Designing nanostructured metamaterials are of great importance to achieve unprecedent multifunction.

The thermal properties of metamaterials can be severely suppressed by designing nanoscale structures, which benefits applications in phononics, thermoelectrics and thermal insulations etc.[11,12] However, most of nanostructured metamaterials are designed relatively simple in morphology, such as nanomesh[5], phononic crystal[7] and nanophononics with local resonators[8]. Designing morphologies of nanostructures provides new degrees of freedom to manipulate thermal properties.[13-16] For example, curved nanostructures, like kinked/bent nanowires and nanoribbon, can cause large modulation of thermal conductivity ($\kappa$).[15,17-21] Moreover, designing film with a wavy-structure can endow rigid film with flexibility.[22,23] Therefore, the thermal properties of metamaterial with designed morphology needs further investigation.

A deformation or strain is inevitable for devices in practical applications. Especially, flexible electronics and devices involve large deformation. Therefore, it is demanded that metamaterials possess stable thermal and mechanical properties under large deformation.[24] However, many investigations found that the deformation has an obvious effects on $\kappa$ of nanostructures.[25-31] Only a few works reported an insensitive $\kappa$ under a smaller deformation (strain <1%.).[32,33] It is less studied that nanostructured metamaterials have insensitive κ under a larger deformation.

Whether a nanostructured metamaterial with designed morphology can have both stable mechanical property and thermal property under deformation? Here, a metamaterial with designed curved Si nanobeams (DCSiNBs) is studied with a large deformation (strain~-41%) by nonequilibrium molecular dynamics (NEMD). The deformation effect on thermal conductance ($\sigma$) is systematically investigated. Further, the stress and phonon density of states (DOS) are calculated to understand the underlying mechanism.

## 2. Design and Method

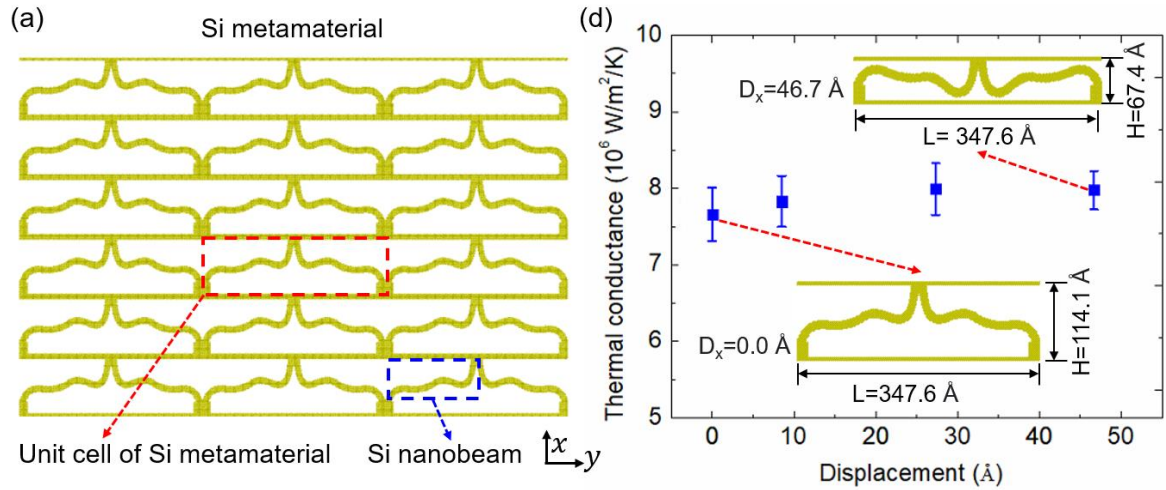

Figure 1. (a) The structure of the designed Si metamaterial. The designed Si metamaterial is constructed by periodic arrangement of the unit cell (shown in the red dashed rectangular). DCSiNBs have a thickness as 10.9 Å (denoted as DCSiNB-I). The generation of DCSiNBs in simulation is shown in Figure S1 in SM. (b) The thermal conductance of the deformed Si metamaterial versus the displacement of top end of the unit cell from 0.0 to 46.7 Å at 300 K in -$x$ direction. The insets show the structure of the unit cell of the designed Si metamaterial without deformation ($D_x$=0.0 Å) and with a deformation ($D_x$=46.7 Å).

The designed Si metamaterial is constructed by periodic arrangement of the unit cell (shown in the red dashed rectangular) in Figure 1(a). The unit cell without deformation has length (L) of 347.6 Å and height (H) of 114.1 Å in $y$ and $x$ direction, respectively. The unit cell is built with DCSiNBs (shown in the blue dashed rectangular). DCSiNBs have a thickness of 10.9 Å (denoted as DCSiNB-I). The deformed Si metamaterial is constructed by the deformed unit cell with deformed DCSiNB. The deformation is described by the displacement ($D_x$) of top end of the unit cell in -$x$ direction. The generation of the DCSiNB with quasi-zero stiffness (QZS) and the procedure to obtain the deformed DCSiNB are shown in Figure S1 in Supplementary Material (SM).

The thermal conductance of the designed Si metamaterial in $x$ direction is calculated by NEMD method using one unit cell by LAMMPS.[34]. The fixed boundary conditions are applied in $x$ direction, and periodic boundary condition is applied in $y$ and $z$ directions. The size of the unit cell in *z* direction is set as 32.6 Å to obtain converged $\sigma$. The interaction between Si atoms are described by Stillinger-Weber potential[35], which has been widely applied in Si nanostructures. Langevin heat baths[36] with temperature of 310 K and 290 K are applied at the two ends of the unit cell in *x* direction, respectively. The time step of NEMD simulation is set as 0.5 fs. In the beginning, the simulation runs 4 ns to reach a steady state. Then, the simulation runs 5 ns to get an averaged heat flux and temperature profile. The thermal conductance is calculated from

$$\sigma = -\frac{J}{A \cdot \Delta T} \tag{1}$$

where $J$ is the total heat current, $A$ is the cross-section area of the unit cell, and $\Delta T$ is the temperature difference between the two ends. The final results of $\sigma$ are averaged over six simulations with different initial conditions. The error bar is the standard deviation of the six simulations.

## 3. Results

The dependence of thermal conductance of the designed Si metamaterial on the displacement of top end of the unit cell from 0.0 to 46.7 Å is shown in Figure 1(b). Interestingly, the $\sigma$ along *x* direction is insensitive to the large deformation (strain~-41%), which is different from the strain effect on Si nanowire and Si film[30]. This result indicates that the designed Si metamaterial can provide a stable thermal property when working under deformation conditions.

The deformation of the designed Si metamaterial is determined by its unit cell and the behavior of DCSiNBs. Therefore, the thermal transport behavior of the deformed DCSiNBs is studied in detail. The deformed DCSiNB-I with $D_x$=0.0, 27.3 and 46.7 Å are shown in Figure 2 (a). In addition, the force-displacement curve of the deformed DCSiNB-I is calculated in Figure S2 in SM, which indicates the QZS feature of DCSiNBs.

The $\sigma$ and temperature profile of the DCSiNB-I in *x* direction is shown in Figure 2 (c) and (b). The structure of the corresponding straight Si beam of DCSiNB-I is shown in Figure S6(a), which has the same thickness and number of atoms as DCSiNB-I. The temperature distributions in both *x* and *y* directions are calculated in Figure S6 in SM. The $\sigma$ of the

DCSiNB-I without deformation ($D_x=0.0$ Å) is reduced 15.8% compared to that of the straight beam without compression (black dashed line), however, it is insensitive to the deformation as the displacement increases, which produces the same trend as that of the designed Si metamaterial.

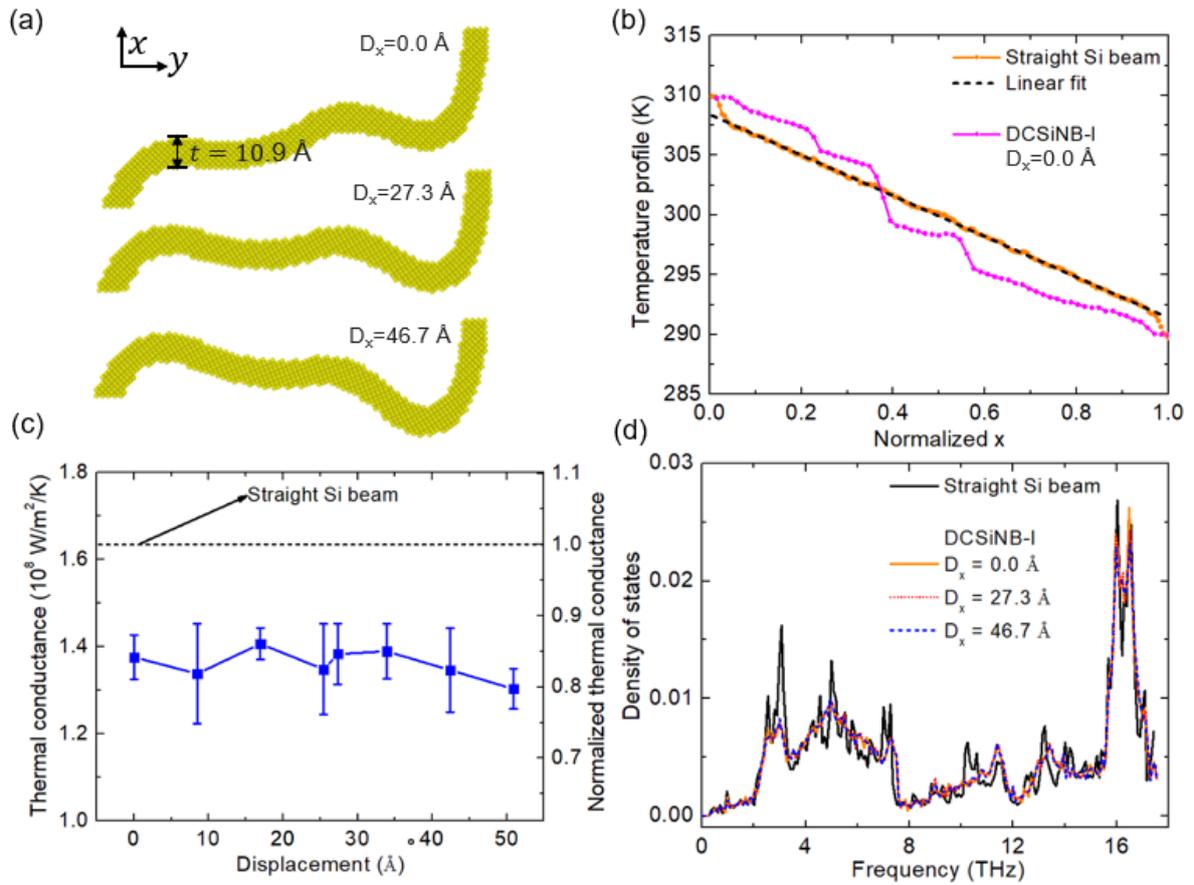

Figure 2. (a) The DCSiNB-I without deformation ($D_x=0.0$) and deformed DCSiNB-I with $D_x=27.3$ and 46.7 Å in -$x$ direction. (b) Temperature profile of the DCSiNB-I with $D_x=0$ Å and the corresponding straight Si beam along $x$ direction. The black dashed line is the linear fit for the straight Si beam. (c) Thermal conductance of the deformed DCSiNB-I with displacement from 0 to 51 Å. The dashed line is for the corresponding straight Si beam without compression. (d) Phonon density of states of straight Si beam and the DCSiNB-I with $D_x=$ 0, 27.32 and 46.67 Å.

To further understand the underlying mechanisms, DOS of the deformed DCSiNB-I with

$D_x$= 0, 27.32 and 46.67 Å and the straight Si beam are calculated by the general utility lattice program (GULP)[37] in Figure 2 (d). The local DOS of atoms in DCSiNB-I are also calculated in Figure S7 in SM. The DOS peaks of the DCSiNB-I are much smaller than that of the straight Si beam when the frequency is between 3.5 and 14 THz. Moreover, the DOS of DCSiNB-I is almost unchanged as the deformation increases, which indicates that the distribution of modes in DCSiNB-I is little affected and can cause the deformation insensitive $\sigma$ of DCSiNB.

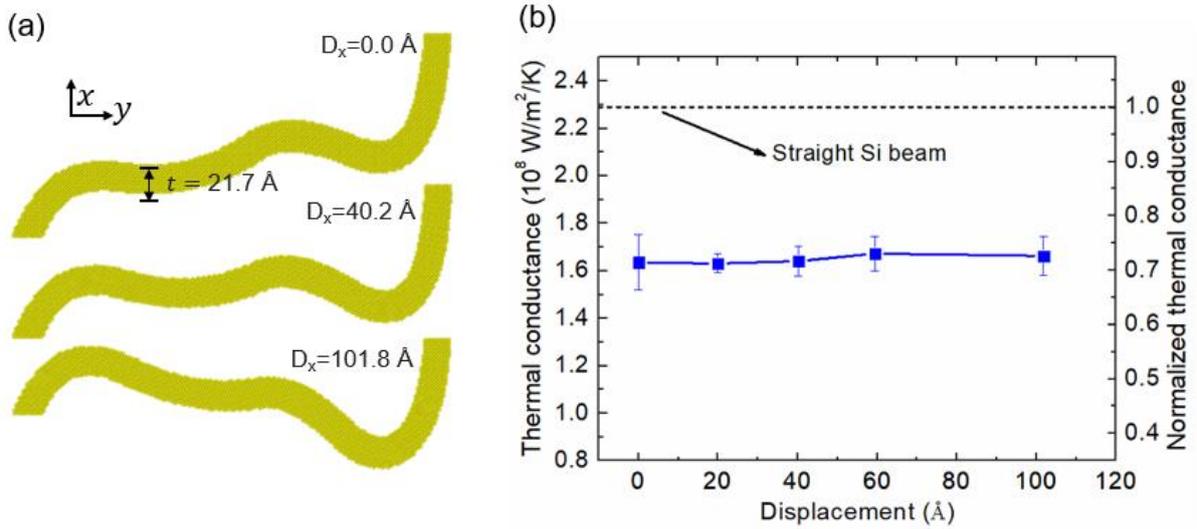

Figure 3. (a) The structure of DCSiNB with thickness of 21.7 Å (denoted as DCSiNB-II). The deformed DCSiNB-II with $D_x$=0.0, 40.2 and 101.8 Å in -$x$ direction are shown. The DCSiNB-II doubles the size of DCSiNB-I in Figure 2 (a). (b) Thermal conductance of the deformed DCSiNB-II versus displacement. The $\sigma$ of the corresponding straight Si beam (black dashed line) is shown for comparison.

To investigate if the size of the DCSiNB can affect the thermal transport behavior under deformation, a thicker DCSiNB (denoted as DCSiNB-II in Figure 3 (a)) whose size doubles that of DCSiNB-I in Figure 2 (a) are studied. The DCSiNB-II also shows a plateau in the force-displacement curve (Figure S2 (a) in SM). The corresponding straight Si beam with thickness of 21.72 Å and length of 47.8 Å is studied for comparison. The temperature profiles and heat flux of DCSiNB-II are calculated in Figure S5 in SM. As shown in Figure 3(b), DCSiNB-II without deformation ($D_x$=0.0 Å) can cause 28.6% reduction of $\sigma$ compared with the straight Si beam. Similar as the DCSiNB-I, the deformed DCSiNB-II also has almost unchanged $\sigma$ as the displacement increases, which further confirms the deformation insensitive $\sigma$ of the designed Si metamaterial.

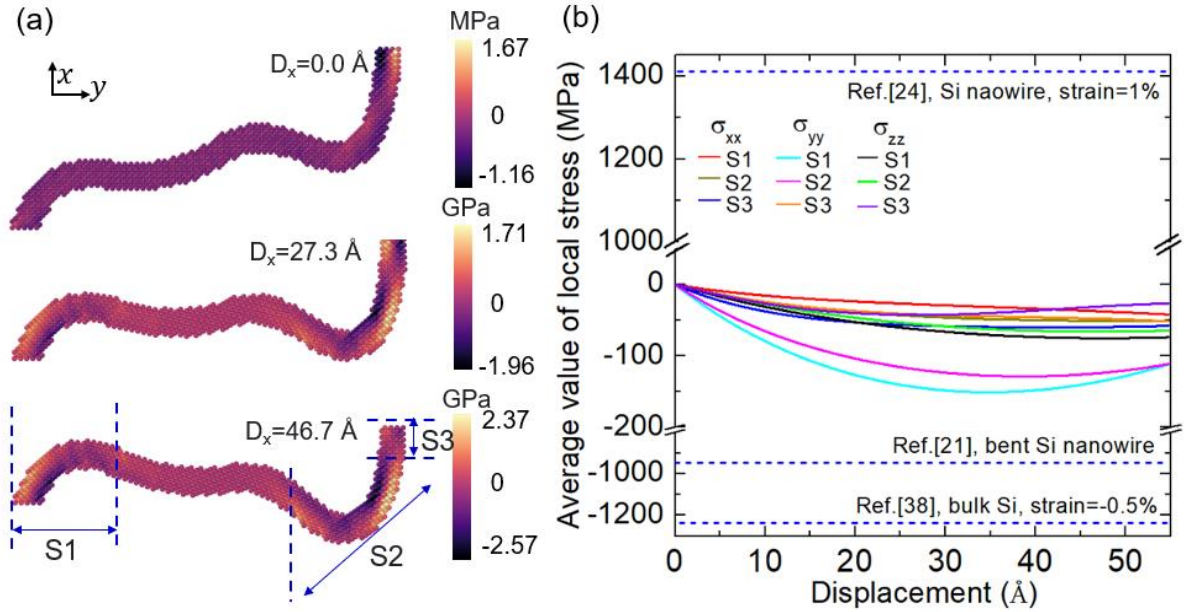

Figure 4. (a) Local stress ($\sigma_{xx}$) distribution in DCSiNB-I without deformation ($D_x$=0 Å) and with deformations ($D_x$= 27.32 and 46.67 Å). The value of the local stress is according the color bars. (b) Average value of local stress in section 1 to 3 (denoted as S1, S2 and S3). The dashed blue lines correspond to the stress in bulk Si with strain=-0.5%[38], bent Si nanowire[21] and Si nanowire with strain=1%[24]. The averaged local stress is ultra-small compared with that in bulk Si and Si nanowire.

To further understand the deformation insensitive $\sigma$, the local stress in the deformed DCSiNB-I with $D_x$=0, 27.32 and 46.67 Å are calculated by LAMMPS[34]. The value of the local stress is according the color bars. Figure 4 (a) shows that only the locations with larger curvature have relatively larger local stress, while most parts have small value of local stress in DCSiNB-I. Further, Figure 4(b) shows that the average values of local stress ($\sigma_{xx}, \sigma_{yy}$ and $\sigma_{zz}$) of the three sections (S1, S2 and S3 in Figure 4(a)) are ultra-small (<151 MPa) compared with the stress of bulk Si with strain=-0.5%[38], Si nanowire with strain=1%[24] and the bent Si nanowire[21] at half diameter region. Consequently, a large deformation just causes small local stress in DCSiNB-I, which in turn leads to the deformation insensitive $\sigma$.

## 4. Conclusion

In this work, the designed Si metamaterial built with DCSiNBs is investigated by NEMD simulations. Interestingly, the $\sigma$ of the designed Si metamaterial is insensitive to large deformation (strain of -41%). The thermal transport behavior of the designed Si metamaterial is determined by DCSiNB which has QZS feature.

Further study confirms that there is an almost unchanged $\sigma$ of DCSiNB under deformation. Under large deformation, the DOS of DCSiNB is little changed and the average value of local stress is ultra-small, which can lead to the deformation insensitive $\sigma$. Besides, compared with the corresponding straight Si beam, the DCSiNB can cause 28.6% reduction of $\sigma$.

The results of this work are meaningful for the multifunctional applications of elaborately design metamaterial with both unchanged thermal conductance and quasi-zero stiffness feature under deformation, such as both stable thermal and stable mechanical properties.

## 5. Acknowledgements

This work is sponsored by the National Natural Science Foundation of China (Grant No. 12004033) (L.Y.), the National Key Research and Development Project of China No. 2018YFE0127800, Fundamental Research Funds for the Central Universities No. 2019kfyRCPY045. The authors thank the National Supercomputing Center in Tianjin (NSCC-TJ) and the China Scientific Computing Grid (ScGrid) for providing assistance in computations.

# Reference

[1] A. Vyatskikh, S. Delalande, A. Kudo, X. Zhang, C.M. Portela, & J.R. Greer, Additive manufacturing of 3D nano-architected metals. Nature communications **9**, 1 (2018).

[2] L.R. Meza, S. Das, & J.R. Greer, Strong, lightweight, and recoverable three-dimensional ceramic nanolattices. Science **345**, 1322 (2014).

[3] J. Bauer, A. Schroer, R. Schwaiger, & O. Kraft, Approaching theoretical strength in glassy carbon nanolattices. Nature materials **15**, 438 (2016).

[4] N.G. Dou, R.A. Jagt, C.M. Portela, J.R. Greer, & A.J. Minnich, Ultralow thermal conductivity and mechanical resilience of architected nanolattices. Nano letters **18**, 4755 (2018).

[5] J.-K. Yu, S. Mitrovic, D. Tham, J. Varghese, & J.R. Heath, Reduction of thermal conductivity in phononic nanomesh structures. Nature Nanotechnology **5**, 718 (2010).

[6] M. Maldovan, Phonon wave interference and thermal bandgap materials. Nature materials **14**, 667 (2015).

[7] L. Yang, N. Yang, & B. Li, Extreme low thermal conductivity in nanoscale 3D Si phononic crystal with spherical pores. Nano Letters **14**, 1734 (2014).

[8] B.L. Davis & M.I. Hussein, Nanophononic metamaterial: Thermal conductivity reduction by local resonance. Physical Review Letters **112**, 055505 (2014).

[9] M.-S. Pham, C. Liu, I. Todd, & J. Lertthanasarn, Damage-tolerant architected materials inspired by crystal microstructure. Nature **565**, 305 (2019).

[10] Q. Zhang, D. Guo, & G. Hu, Tailored Mechanical Metamaterials with Programmable Quasi-Zero-Stiffness Features for Full-Band Vibration Isolation. Advanced Functional Materials **31**, 2101428 (2021).

[11] L. Yang, D. Huh, R. Ning, V. Rapp, Y. Zeng, Y. Liu, S. Ju, Y. Tao, Y. Jiang, & J. Beak, High thermoelectric figure of merit of porous Si nanowires from 300 to 700 K. Nature Communications **12**, 1 (2021).

[12] X. Qian, J. Zhou, & G. Chen, Phonon-engineered extreme thermal conductivity materials. Nature Materials, 1 (2021).

[13] C. Portela & J. Ye, Architectures down to nano. Nature Materials **20**, 1451 (2021).

[14] B. Tian, P. Xie, T.J. Kempa, D.C. Bell, & C.M. Lieber, Single-crystalline kinked

semiconductor nanowire superstructures. Nature Nanotechnology **4**, 824 (2009).

[15] Q. Zhang, Z. Cui, Z. Wei, S.Y. Chang, L. Yang, Y. Zhao, Y. Yang, Z. Guan, Y. Jiang, & J. Fowlkes, Defect facilitated phonon transport through kinks in boron carbide nanowires. Nano Letters **17**, 3550 (2017).

[16] L. Yang, R. Prasher, & D. Li, From nanowires to super heat conductors. Journal of Applied Physics **130**, 220901 (2021).

[17] J.-W. Jiang, N. Yang, B.-S. Wang, & T. Rabczuk, Modulation of thermal conductivity in kinked silicon nanowires: phonon interchanging and pinching effects. Nano Letters **13**, 1670 (2013).

[18] Y. Zhao, L. Yang, C. Liu, Q. Zhang, Y. Chen, J. Yang, & D. Li, Kink effects on thermal transport in silicon nanowires. International Journal of Heat and Mass Transfer **137**, 573 (2019).

[19] L. Yang, Q. Zhang, Z. Wei, Z. Cui, Y. Zhao, T.T. Xu, J. Yang, & D. Li, Kink as a new degree of freedom to tune the thermal conductivity of Si nanoribbons. Journal of Applied Physics **126**, 155103 (2019).

[20] L.-C. Liu, M.-J. Huang, R. Yang, M.-S. Jeng, & C.-C. Yang, Curvature effect on the phonon thermal conductivity of dielectric nanowires. Journal of Applied Physics **105**, 104313 (2009).

[21] X. Liu, H. Zhou, G. Zhang, & Y.-W. Zhang, The effects of curvature on the thermal conduction of bent silicon nanowire. Journal of Applied Physics **125**, 082505 (2019).

[22] Y. Liu, K. He, G. Chen, W.R. Leow, & X. Chen, Nature-inspired structural materials for flexible electronic devices. Chemical Reviews **117**, 12893 (2017).

[23] D.-Y. Khang, H. Jiang, Y. Huang, & J.A. Rogers, A stretchable form of single-crystal silicon for high-performance electronics on rubber substrates. Science **311**, 208 (2006).

[24] H. Zhang, J. Tersoff, S. Xu, H. Chen, Q. Zhang, K. Zhang, Y. Yang, C.-S. Lee, K.-N. Tu, & J. Li, Approaching the ideal elastic strain limit in silicon nanowires. Science Advances **2**, e1501382 (2016).

[25] A.R. Abramson, C.-L. Tien, & A. Majumdar, Interface and strain effects on the thermal conductivity of heterostructures: A molecular dynamics study. Journal of Heat Transfer **124**, 963 (2002).

[26] Y. Kuang, L. Lindsay, S. Shi, X. Wang, & B. Huang, Thermal conductivity of graphene mediated by strain and size. International Journal of Heat and Mass Transfer **101**, 772 (2016).

[27] Q. Wang, L. Han, L. Wu, T. Zhang, S. Li, & P. Lu, Strain effect on thermoelectric performance of InSe monolayer. Nanoscale Research Letters **14**, 1 (2019).

[28] B. Ding, X. Li, W. Zhou, G. Zhang, & H. Gao, Anomalous strain effect on the thermal conductivity of low-buckled two-dimensional silicene. National Science Review **8**, nwaa220 (2021).

[29] Z. Yang, R. Feng, F. Su, D. Hu, X. Ma, & Nanostructures, Isotope and strain effects on thermal conductivity of silicon thin film. Physica E: Low-dimensional Systems **64**, 204 (2014).

[30] X. Li, K. Maute, M.L. Dunn, & R. Yang, Strain effects on the thermal conductivity of nanostructures. Physical Review B **81**, 245318 (2010).

[31] M. Alam, M.P. Manoharan, M.A. Haque, C. Muratore, A. Voevodin, & Microengineering, Influence of strain on thermal conductivity of silicon nitride thin films. Journal of Micromechanics Microengineering **22**, 045001 (2012).

[32] D. Fan, H. Sigg, R. Spolenak, & Y. Ekinci, Strain and thermal conductivity in ultrathin suspended silicon nanowires. Physical Review B **96**, 115307 (2017).

[33] K.F. Murphy, B. Piccione, M.B. Zanjani, J.R. Lukes, & D.S. Gianola, Strain-and defect-mediated thermal conductivity in silicon nanowires. Nano Letters **14**, 3785 (2014).

[34] A.P. Thompson, H.M. Aktulga, R. Berger, D.S. Bolintineanu, W.M. Brown, P.S. Crozier, P.J. in't Veld, A. Kohlmeyer, S.G. Moore, & T.D. Nguyen, LAMMPS-a flexible simulation tool for particle-based materials modeling at the atomic, meso, and continuum scales. Computer Physics Communications **271**, 108171 (2022).

[35] F.H. Stillinger & T.A. Weber, Computer simulation of local order in condensed phases of silicon. Physical Review B **31**, 5262 (1985).

[36] J. Chen, G. Zhang, & B. Li, Molecular dynamics simulations of heat conduction in nanostructures: effect of heat bath. Journal of the Physical Society of Japan **79**, 074604 (2010).

[37] J.D. Gale & A.L. Rohl, The general utility lattice program (GULP). Molecular Simulation

**29**, 291 (2003).

[38] V. Kuryliuk, O. Nepochatyi, P. Chantrenne, D. Lacroix, & M. Isaiev, Thermal conductivity of strained silicon: Molecular dynamics insight and kinetic theory approach. Journal of Applied Physics **126**, 055109 (2019).